\documentstyle[amsfonts,aps]{revtex}

\begin{document}
\title{Functional Approach to Quantum Decoherence and the Classical Final Limit.}
\author{Mario Castagnino}
\address{Instituto de Astronom\'{\i}a y F\'{\i}sica del Espacio.\\
Casilla de Correos 67, Sucursal 28\\
1428, Buenos Aires, Argentina.}
\author{Roberto Laura}
\address{Departamento de F\'{\i}sica, F.C.E.I.A, Universidad Nacional de Rosario\\
Instituto de F\'{\i}sica de Rosario, CONICET-UNR\\
Av. Pellegrini 250, 2000 Rosario, Argentina.}
\date{November 1999}
\maketitle

\begin{abstract}
For a wide set of quantum systems it is demonstrated that the quantum regime
can be considered as the transient phase while the final classical
statistical regime is a permanent state. A basis where exact matrix
decoherence appears for these final states is found. The relation with the
decoherence of histories formalism is studied. A set of final intrinsically
consistent histories is found.
\end{abstract}

\section{Introduction.}

Following the idea that the interplay of observables and states is the
fundamental ingredient of quantum mechanics \footnote{%
According to W. Zurek:...''the only sensible subject of consideration aimed
at the interpretation of quantum theory... is the {\it relation} between the
universal {\it state vector} and the state memory (records) of somewhat
special system - such as {\it observers }- which are, for necessity,
perceiving the Universe from within. It is the inability to appreciate the
consequences of this rather simple but fundamental observation that has led
to such desperate measures as the search of an alternative quantum physics 
\cite{Zurek}''.} we have developed paper \cite{CyLII} where we have studied
the {\it relation} of the {\it state vectors }$\rho $ of a close isolated
quantum system (that belong to a convex set of states ${\cal S}$), to the 
{\it observables }$O$ within this closed system (that belong to a space of
observables ${\cal O}$). We consider that the essence of this relation is
the mean value of an observable $O$ in a state $\rho ,$ which is given by
the equation: 
\begin{equation}
\langle O\rangle _\rho =Tr(\rho O)=(\rho |O)  \label{0.1}
\end{equation}
In fact at the statistical level what we actually measure in an ensemble of
identical states, are these kind of averages, since we cannot either measure
directly the state $\rho $ or measure it with an infinite precision \cite
{Ballentine}. Moreover, these averages can be considered, as in the r. h. s.
of eq. (\ref{0.1}), the result of a linear functional $(\rho |\in {\cal S}$
acting on a vector $|O)\in {\cal O,}$ and therefore we can say that ${\cal %
S\subset O}^{^{\prime }}$, being ${\cal O}^{^{\prime }}$ the dual of space $%
{\cal O}$. While for the usual states (mixed or pure) we can use $Tr(\rho O)$%
, there are {\it generalized states} that can be defined as the functional $%
(\rho |O)$ as explained in papers \cite{CyLII}. Many results were obtained
using this formalism (see e. g.: \cite{CyLII}, \cite{LyCE}, \cite{Antoniou}).

In this paper we will use the formalism of paper \cite{CyLII} to study the
so called ''classical limit problem'', namely the {\it statistical quantum
mechanics }$\rightarrow $ {\it classical mechanics} limit that appears in
some quantum systems when observed using certain spaces of observables $%
{\cal O}$ . For conceptual reasons we will divide the problem in two
different processes (that may or may not happen simultaneously):

(a)-{\it \ Statistical process: }Namely the limit {\it statistical quantum
mechanics} $\rightarrow $ {\it statistical classical mechanics, }where the
phenomenon of {\it decoherence} combined with the disappearance of the
uncertainty relations in the limit $\hbar \rightarrow 0,$ originates the
classical final stationary state. Almost all the paper will be devoted to
this problem . We will see how when $t\rightarrow \infty $ the quantum
system reaches a classical final stationary state $\rho _{*}(q,p)$ where the
statistical dynamics is trivial, since $\rho _{*}(q,p)$ is time independent,
but the systems of the ensemble move according to the non-trivial laws of
classical dynamics. In general we will have an unlocalized statistical
classical state of many identical systems moving in phase space.

(b)-{\it \ Localization process: } It is the evolution {\it statistical
classical mechanics }$\rightarrow $ {\it classical mechanics. }In some
special cases the evolution privileges a single space-time trajectory, in
such a way that all trajectories (endowed with a non negligible positive
probability) concentrate around it\footnote{%
In some cases this phenomenon does not happen for all the systems but only
for a subsystem.}. In this case we will have {\it correlations }and {\it %
localization. }Then we have the statistical classical state of all the
systems practically moving along the same trajectory in such a way that we
may consider that we are dealing with a{\it \ single classical system. }We
will discuss this process in section IV and appendix B.

The usual technique to solve these problems is coarse-graining. But in our
method we will consider not only the coarse-graining average but{\it \ all}
possible averages made using the observables of space ${\cal O}$, thus we
are generalizing the coarse-graining idea \footnote{%
At least the ''coarse graining alla Zurek''.}. In fact, among the
observables of ${\cal O}$ there are some that, from the density matrix $\rho
,$ take into account only some component $\rho _r$, the so called relevant
part of $\rho ,$ and completely neglect the complementary component $\rho
_i, $ the so called irrelevant part of $\rho ,$ i. e. these observables only
measure (macroscopic) properties of what it is considered as the ''system''
(contained in $\rho _r)$ and neglect or average the (microscopic) properties
of the ''environment'' (contained in $\rho _i).$ But we will consider not
only this kind of observables but all observables in ${\cal O.}$ Therefore
the interplaying of observables and states will take the role of the
coarse-graining in this paper (see also the end of section IIA). With this
strategy we cannot only obtain all the old results, but also we will find
some new ones.

We will use this method to study the process (a) and to prove that certain
quantum systems evolve from a statistical quantum state to the statistical
classical final stationary state. In the same framework we will study the
process (b) obtaining the classical motion of a single system.

The paper is organized as follows:

In section II we will see, using the Riemann-Lebesgue theorem, that
transition (a) takes place in close systems endowed with a continuous
spectrum and with just one bound state (as in the classical mixing systems).
More general cases will be considered in section II C. The main
characteristics of the quantum laws are:

1.- The non-boolean nature of the way to find the probability of two
exclusive events (this probability is the square modulus of the sum of their
amplitudes and not the sum of the probabilities).

2.- The uncertainty relations.

In the evolution from quantum mechanics to classical mechanics the first
characteristic disappears (and the boolean way of adding probabilities is
established) by the process of {\it decoherence} and the uncertainty
relations can be neglected in the limit $\hbar \rightarrow 0.$ Then we can
use the laws of {\it classical statistical mechanics.}

At this stage four remarks are in order:

i.- Using our language the generalized idea of decoherence can be introduced
in the following way: At the quantum level the average (\ref{0.1}) reads: 
\begin{equation}
\langle O\rangle _\rho ^{(q)}=\sum_{\omega ,\omega ^{\prime }}\rho _{\omega
\omega ^{\prime }}O_{\omega ^{\prime }\omega }  \label{0.2}
\end{equation}
where $\rho _{\omega \omega ^{\prime }}$ and $O_{\omega \omega ^{\prime }}$
are the components in some basis of the operators $\rho $ and $O$
respectively. Eq. (\ref{0.2}) can be considered as the average of some
quantities $O_{\omega \omega ^{\prime }}$ weighted by some generalized
correlations $\rho _{\omega \omega ^{\prime }}$ (since the $\rho _{\omega
\omega }$ are probabilities but the $\rho _{\omega \omega ^{\prime }}$, with 
$\omega \neq \omega ^{\prime },$ are quantum correlations). On the other
hand, at the classical level we also have some quantities $O_\omega $ that
correspond to a set $\{\omega \}$ of the exhaustive and exclusive
alternatives, each one with a (boolean) probability $p_\omega $ of measure $%
\omega $ for the observable $O.$ The corresponding classical weighted
average is: 
\begin{equation}
\langle O\rangle _\rho ^{(cl)}=\sum_\omega p_\omega O_\omega  \label{0.3}
\end{equation}
where $\sum_\omega p_\omega =1.$ The transition from the quantum phase to
the classical one is therefore: 
\begin{equation}
\sum_{\omega ,\omega ^{\prime }}\rho _{\omega \omega ^{\prime }}O_{\omega
^{\prime }\omega }\rightarrow \sum_\omega p_\omega O_\omega  \label{0.4}
\end{equation}
at least for some $O$ which belong to a preferred sub space of ${\cal O}$
(i.e. to a subspace expanded by a complete set of commuting observables, a
CSCO, that we will define below; the eigenbasis of this set will be the so
called {\it final pointer basis}). If in (\ref{0.4}) we take $\rho _{\omega
\omega }=p_\omega $ and $O_{\omega \omega }=O_\omega $, the matrix $\rho
_{\omega \omega ^{\prime }}$ must become diagonal in the final pointer
basis. This is the essence of the transition (a), since the above relation
will be valid for all observables of the CSCO and we will have: 
\begin{equation}
\langle O\rangle _\rho ^{(q)}\rightarrow \langle O\rangle _\rho ^{(cl)}
\label{0.5}
\end{equation}
If this transition takes place, boolean logic is established in the
statistical classical system, if we perform the measurement with the
observables of the preferred CSCO. In the usual parlance we will then say
that the density matrices that contain quantum interference terms become
diagonal, in such away that these interferences are suppressed. Then the
quantum way to find probabilities of exclusive and exhaustive alternatives,
i. e.: adding the corresponding amplitudes and computing the norm, becomes
the classical boolean way: just adding the probabilities.

ii.- In this paper decoherence is essentially studied in systems with
continuous spectrum. The case of the discrete spectrum, and the causes of
decoherence in this case, are discussed in section II C.

iii.- In the case of the continuous spectrum the essence of the method is
the following: If $\omega \in {\Bbb R}^{+}$ are the eigenvalues of $H$ and
we call $\nu =\omega -\omega ^{\prime },$ the $\rho _{\omega \omega ^{\prime
}}$ of eq. (\ref{0.2}) is a function $\rho (\nu ,...).$ Then the time limit
of its evolution is given by the Riemann-Lebesgue theorem, that prescribes
that: 
\begin{equation}
\lim_{t\rightarrow \infty }\int_{-a}^ae^{-i\nu t}\rho (\nu ,...)d\nu =0
\label{0.5'}
\end{equation}
if $\rho (\nu ,...)$ is integrable. So all the diagonal terms ($\nu =0)$ and
all the off-diagonal terms ($\nu \neq 0)$ vanish. Therefore this theorem
cannot be used as a computation method in the case of continuous spectrum.
Nevertheless when we consider the problem within a cube of size $L$, we
define $\rho _{\omega \omega ^{\prime }}$ there, and when we make $%
L\rightarrow \infty $, it can be shown that a s{\it ingular structure }%
appears for $\rho (\nu ,...)$ and the corresponding singular diagonal term
remains as it should. The method introduced in paper \cite{CyLII} is
precisely designed to rigorously deal with these singular structures. It has
yielded good results in papers \cite{CyLII}, \cite{Antoniou}, \cite{LyCE}.

iv.- Before the classical stationary state limit is reached usually the
system goes through a ''classical phase'' where the state can be considered
as classical but it is not yet in its final classical stationary state. But
our method can only be used when $t\rightarrow \infty $. It only allows to
find the ''statistical classical final limit''. So, we essentially study
this final stationary state but we believe that our method can be
generalized to cover the classical phase before the final stationary state,
so we will discuss these matters in section V. Moreover, we believe that the
understanding of the final limit will enhance the chances to understand the
much more difficult problem of the classical phase, in the clearest and
concise way.

In section III we reach to the principal aim of the formalism of transition
''a'' which is to create a bridge between quantum and classical mechanics,
precisely between quantum mechanics and classical {\it statistical}
mechanics at equilibrium. We know that the uncertainty relations disappear,
when $\hbar \rightarrow 0$, (more precisely when the characteristic
dimension of the system makes $\hbar $ a negligible quantity). Then, let us
consider a system where the quantum state is defined by a density matrix $%
\rho ,$ and a set of classical trajectories in phase space labelled by some
constants $x,l_1,...,l_N$, $a_1,...,a_N,$ where $x$ corresponds to the
energy, $l_1,...,l_N$ to other dynamical momentum variables, and, $%
a_1,...,a_N$ to configuration variables. The aim of the theory is:

1.- To transform the matrix $\rho $ into a classical density function in
phase space $\rho (q,p)$ when $\hbar \rightarrow 0.$

2.- To decompose $\rho (q,p)$ as: 
\begin{equation}
\rho (q,p)=\sum_{x,l_1,...,l_N,a_1,...,a_N}p_{x,l_1,...,l_N,a_1,...,a_N}\rho
_{x,l_1,...,l_N,a_1,...,a_N}(q,p)  \label{0.6}
\end{equation}
where $q$ and $p$ are the position and momentum coordinates and the
classical densities $\rho _{x,l_1,...,l_N,a_1,...,a_N}(q,p)$ would
correspond to each classical trajectory \footnote{%
The dimension of the phase space considered is $2(N+1)$. Then there are $%
(N+1)$ momenta and ($N+1)$ coordinates. So $N$ $+1$ is the number of
parameters necessary to label the momenta of the classical space-time
trajectories, and $N$ the number necessary to label the origins of the
trajectories.} ( in the classical sense that it is peaked in the trajectory
and thus it rapidly vanishes when going from the near vicinity of the
trajectory to the far zones of the phase space) and $%
p_{x,l_1,...,l_N,a_1,...,a_N}$ is the probability of each trajectory.

We will obtain (when $\hbar \rightarrow 0$) these results as follows:

1.-$\rho (q,p)$ will be the Wigner function corresponding to the matrix $%
\rho .$

2.-$\rho _{x,l_1,...,l_N,a_1,...,a_N}(q,p)$ will be the Wigner functions of
the wave packets going along the classical trajectories labelled by the
constant of the motion $x,l_1,...,l_N,$ and passing by the initial point of
coordinates $a_1,...,a_N.$

We will see that all this happens after a convenient decoherence time and we
will obtain the last expansion (cf. eq. (\ref{5.15})) and therefore what we
consider the best bridge between classical and quantum statistical concepts
(see paper \cite{Zoupas} for a very similar conclusion).

We will devote section IV to discuss transition (b), namely the localization
process. Eventually in some cases this process takes place and {\it %
correlations} appear and we reach to a single classical state if the
localization process is efficient enough. Then we can use the laws of {\it %
classical mechanics. }This phenomenon happens if the dynamic of the system
and the initial conditions are such that some canonically conjugated
variables correlate (see appendix B). We will see how this fact can be
incorporated in our formalism.

We will draw our main conclusions and comments in section V.

Appendix A is devoted to compare our results with those in the literature.
In appendix B we deal with correlations and localization. Finally, in
appendix C we translate the results into the language of usual decoherence
of histories.

\section{Decoherence.}

\subsection{Decoherence in the energy.}

Let us consider an isolated quantum system with $N+1$ dynamical variables
and a Hamiltonian endowed with a continuous spectrum and just one bounded
state. So the discrete part of the spectrum of $H$ has only one value $%
\omega _0$ and the continuous spectrum is let say $0\leq \omega <\infty $
(how the discrete spectrum behaves in the continuous limit can be seen in
papers \cite{Alimanias}, \cite{Gruver}). Eventually we will give the
collective name $x$ to both $\omega _0$ and $\omega .$ Let us assume that it
is possible to diagonalize the Hamiltonian $H$, together with $N$
observables $O_i$ ($i=1,...,N)$. The operators ($H$, $O_1$,...,$O_N$) form a 
{\it complete set of commuting observables} (CSCO). For simplicity we also
assume a discrete spectrum for the $N$ observables $O_i$. Therefore we write 
\begin{equation}
H=\omega _0\sum_m|\omega _0,m\rangle \langle \omega _0,m|+\int_0^\infty
\omega \sum_m|\omega ,m\rangle \langle \omega ,m|d\omega  \label{2.4}
\end{equation}
where $\omega _0<0$ is the energy of the ground state, and $m\doteq
\{m_1,...,m_N\}$ labels a set of discrete indexes which are the eigenvalues
of the observables $O_1$,...,$O_N$. $\{|\omega _0,m\rangle ,|\omega
,m\rangle \}$ is a basis of simultaneous generalized eigenvectors of the
CSCO: 
\[
H|\omega _0,m\rangle =\omega _0|\omega _0,m\rangle ,\quad H|\omega ,m\rangle
=\omega |\omega ,m\rangle ,\quad 
\]

\[
O_i|\omega _0,m\rangle =m_i|\omega _0,m\rangle ,\quad O_i|\omega ,m\rangle
=m_i|\omega ,m\rangle . 
\]
The most general observable that we are going to consider in our model
reads: 
\begin{eqnarray}
O &=&\sum_{mm^{\prime }}O(\omega _0)_{mm^{\prime }}|\omega _0,m\rangle
\langle \omega _0,m^{\prime }|+\sum_{mm^{\prime }}\int_0^\infty d\omega
O(\omega )_{mm^{\prime }}|\omega ,m\rangle \langle \omega ,m^{\prime
}|+\sum_{mm^{\prime }}\int_0^\infty d\omega O(\omega ,\omega _0)_{mm^{\prime
}}|\omega ,m\rangle \langle \omega _0,m^{\prime }|+  \nonumber \\
&&+\sum_{mm^{\prime }}\int_0^\infty d\omega ^{\prime }O(\omega _0,\omega
^{\prime })_{mm^{\prime }}|\omega _0,m\rangle \langle \omega ^{\prime
},m^{\prime }|+\sum_{mm^{\prime }}\int_0^\infty \int_0^\infty d\omega
d\omega ^{\prime }O(\omega ,\omega ^{\prime })_{mm^{\prime }}|\omega
,m\rangle \langle \omega ^{\prime },m^{\prime }|,  \label{2.5}
\end{eqnarray}
where $O(\omega )_{mm^{\prime }}$, $O(\omega ,\omega _0)_{mm^{\prime }}$, $%
O(\omega _0,\omega )_{mm^{\prime }}$ and $O(\omega ,\omega ^{\prime
})_{mm^{\prime }}$ are ordinary functions of the real variables $\omega $
and $\omega ^{\prime }$(these functions must have some mathematical
properties in order to develop the theory; these properties are listed in
paper \cite{CyLII}). Namely, the most general observables have a singular
component (the second term of the r.h.s. of the last equation) and a regular
part (all the other terms). If the singular term would be missing the
Hamiltonian (\ref{2.4}) would not belong to the space of the chosen
observables \cite{CyLII}. We will say that these observables belong to a
space ${\cal O}$. This space has the {\it basis} $\{|\omega _0,mm^{\prime })$%
, $|\omega ,mm^{\prime })$, $|\omega \omega _0,mm^{\prime })$, $|\omega
_0\omega ^{\prime },mm^{\prime })$, $|\omega \omega ^{\prime },mm^{\prime
})\}$: 
\[
|\omega _0,mm^{\prime })\doteq |\omega _0,m\rangle \langle \omega
_0,m^{\prime }|,\quad \quad |\omega ,mm^{\prime })\doteq |\omega ,m\rangle
\langle \omega ,m^{\prime }|,\quad \quad |\omega \omega _0,mm^{\prime
})\doteq |\omega ,m\rangle \langle \omega _0,m^{\prime }|, 
\]
\begin{equation}
|\omega _0\omega ^{\prime },mm^{\prime })\doteq |\omega _0,m\rangle \langle
\omega ^{\prime },m^{\prime }|,\quad \quad |\omega \omega ^{\prime
},mm^{\prime })\doteq |\omega ,m\rangle \langle \omega ^{\prime },m^{\prime
}|.  \label{2.5'}
\end{equation}
The quantum states $\rho $ are measured by the observables just defined,
computing the mean values of these observables in the quantum states, i. e.
in the usual notation: $\langle O\rangle _\rho =Tr(\rho ^{\dagger }O)$ \cite
{Ballentine}. These mean values, generalized as in paper \cite{CyLII}, can
be considered as linear functionals $\rho ,$ mapping the vectors $O$ on the
real numbers$,$ that we can call $(\rho |O)$ \cite{Bogo}. In fact, this is a
generalization of the usual mean value definition. Then $\rho \in {\cal %
S\subset O}^{^{\prime }},$ where ${\cal S}$ is a convenient convex set
contained in ${\cal O}^{^{\prime }}$, the space of linear functionals over $%
{\cal O}$ \cite{CyLI}, \cite{CyLIII}. The basis of ${\cal O}^{\prime }$
(that can also be considered as the {\it co-basis} of ${\cal O)}$ is $%
\{(\omega _0,mm^{\prime }|$, $(\omega ,mm^{\prime }|$, $(\omega \omega
_0,mm^{\prime }|$, $(\omega _0\omega ^{\prime },mm^{\prime }|$, $(\omega
\omega ^{\prime },mm^{\prime }|\}$ defined as functionals by the equations: 
\[
(\omega _0,mm^{\prime }|\omega _0,nn^{\prime })=\delta _{mn}\delta
_{m^{\prime }n^{\prime }},\quad (\omega ,mm^{\prime }|\eta ,nn^{\prime
})=\delta (\omega -\eta )\delta _{mn}\delta _{m^{\prime }n^{\prime }},\quad
(\omega \omega _0,mm^{\prime }|\eta \omega _0,nn^{\prime })=\delta (\omega
-\eta )\delta _{mn}\delta _{m^{\prime }n^{\prime }}, 
\]
\begin{equation}
(\omega _0\omega ^{\prime },mm^{\prime }|\omega _0\eta ^{\prime },nn^{\prime
})=\delta (\omega ^{\prime }-\eta ^{\prime })\delta _{mn}\delta _{m^{\prime
}n^{\prime }},\quad (\omega \omega ^{\prime },mm^{\prime }|\eta \eta
^{\prime },nn^{\prime })=\delta (\omega -\eta )\delta (\omega ^{\prime
}-\eta ^{\prime })\delta _{mn}\delta _{m^{\prime }n^{\prime }}.
\label{2.5''}
\end{equation}
and all other $(.|.)$ are zero. In particular we have 
\begin{equation}
(\omega _0,mm^{\prime }|O)=O(\omega _0)_{mm^{\prime }}=\langle \omega
_0,m|O|\omega _0,m^{\prime }\rangle  \label{2.5A}
\end{equation}
for any $O\in {\cal O}$. But $(\omega ,mm^{\prime }|O)=O(\omega
)_{mm^{\prime }}$ is not equal to $\langle \omega ,m|O|\omega ,m^{\prime
}\rangle $, which is not even defined if $O$ is given by eq.(\ref{2.5}).
Therefore $(\omega ,mm^{\prime }|$ can only be considered as a functional,
being a typical generalized state. Then, a generic quantum state reads: 
\begin{eqnarray}
\rho &=&\sum_{mm^{\prime }}\overline{\rho (\omega _0)}_{mm^{\prime }}(\omega
_0,mm^{\prime }|+\sum_{mm^{\prime }}\int_0^\infty d\omega \overline{\rho
(\omega )}_{mm^{\prime }}(\omega ,mm^{\prime }|++\sum_{mm^{\prime
}}\int_0^\infty d\omega \overline{\rho (\omega ,\omega _0)}_{mm^{\prime
}}(\omega \omega _0,mm^{\prime }|+  \nonumber \\
&&+\sum_{mm^{\prime }}\int_0^\infty d\omega ^{\prime }\overline{\rho (\omega
_0,\omega ^{\prime })}_{mm^{\prime }}(\omega _0\omega ^{\prime },mm^{\prime
}|+\sum_{mm^{\prime }}\int_0^\infty d\omega \int_0^\infty d\omega ^{\prime }%
\overline{\rho (\omega ,\omega ^{\prime })}_{mm^{\prime }}(\omega \omega
^{\prime },mm^{\prime }|,  \label{2.6}
\end{eqnarray}
where $\overline{\rho (\omega _0)}_{mm}$ and $\overline{\rho (\omega )}_{mm}$
are real and non negative, $\overline{\rho (\omega ,\omega _0)}_{mm^{\prime
}}=\rho (\omega _0,\omega )_{m^{\prime }m}$ and $\overline{\rho (\omega
,\omega ^{\prime })}_{mm^{\prime }}=\rho (\omega ^{\prime },\omega
)_{m^{\prime }m}$. Moreover, $\rho (\omega _0)_{mm^{\prime }}$ and $\rho
(\omega )_{mm^{\prime }}$ satisfy the total probability condition 
\begin{equation}
(\rho |I)=\sum_m\rho (\omega _0)_{mm}+\sum_m\int_0^\infty d\omega \rho
(\omega )_{mm}=1,  \label{2.6'}
\end{equation}
where $I=\sum_m|\omega _0,m\rangle \langle \omega _0,m|+\int_0^\infty
d\omega \sum_m|\omega ,m\rangle \langle \omega ,m|$ is the identity operator
in ${\cal O}$. Eq. (\ref{2.6'}) is the extension to state functionals of the
usual condition $Tr\rho ^{\dagger }=1$, used when $\rho $ is a density
operator.

The time evolution of the quantum state $\rho $ reads: 
\begin{eqnarray}
\rho (t) &=&\sum_{mm^{\prime }}\overline{\rho (\omega _0)}_{mm^{\prime
}}(\omega _0,mm^{\prime }|+\sum_{mm^{\prime }}\int_0^\infty d\omega 
\overline{\rho (\omega )}_{mm^{\prime }}(\omega ,mm^{\prime
}|+\sum_{mm^{\prime }}\int_0^\infty d\omega \overline{\rho (\omega ,\omega
_0)}_{mm^{\prime }}e^{i(\omega -\omega _0)t}(\omega \omega _0,mm^{\prime }|+
\nonumber \\
&&+\sum_{mm^{\prime }}\int_0^\infty d\omega ^{\prime }\overline{\rho (\omega
_0,\omega ^{\prime })}_{mm^{\prime }}e^{i(\omega _0-\omega ^{\prime
})t}(\omega _0\omega ^{\prime },mm^{\prime }|+\sum_{mm^{\prime
}}\int_0^\infty d\omega \int_0^\infty d\omega ^{\prime }\overline{\rho
(\omega ,\omega ^{\prime })}_{mm^{\prime }}e^{i(\omega -\omega ^{\prime
})t}(\omega \omega ^{\prime },mm^{\prime }|  \label{2.7}
\end{eqnarray}

The mean value of an observable $O$ in a quantum state $\rho $ reads:

\begin{eqnarray}
\langle O\rangle _{\rho (t)} &=&(\rho (t)|O)=  \nonumber \\
&=&\sum_{mm^{\prime }}\overline{\rho (\omega _0)_{mm^{\prime }}}O(\omega
_0)_{mm^{\prime }}+\sum_{mm^{\prime }}\int_0^\infty d\omega \overline{\rho
(\omega ,\omega ^{\prime }})_{mm^{\prime }}O(\omega )_{mm^{\prime }}+ 
\nonumber \\
&&+\sum_{mm^{\prime }}\int_0^\infty d\omega \overline{\rho (\omega ,\omega
_0)}_{mm^{\prime }}e^{i(\omega -\omega _0)t}O(\omega ,\omega _0)_{mm^{\prime
}}+\sum_{mm^{\prime }}\int_0^\infty d\omega ^{\prime }\overline{\rho (\omega
_0,\omega ^{\prime })}_{mm^{\prime }}e^{i(\omega _0-\omega ^{\prime
})t}O(\omega _0,\omega ^{\prime })_{mm^{\prime }}  \nonumber \\
&&+\sum_{mm^{\prime }}\int_0^\infty d\omega \int_0^\infty d\omega ^{\prime }%
\overline{\rho (\omega ,\omega ^{\prime })}_{mm^{\prime }}e^{i(\omega
-\omega ^{\prime })t}O(\omega ,\omega ^{\prime })_{mm^{\prime }}.
\label{2.8}
\end{eqnarray}
Using the Riemann-Lebesgue theorem we obtain the weak limit, for all $O\in 
{\cal O}$ 
\begin{equation}
\lim_{t\rightarrow \infty }\langle O\rangle _{\rho (t)}=\langle O\rangle
_{\rho _{*}}  \label{2.9}
\end{equation}
where we have introduced the diagonal asymptotic or final stationary state
functional 
\begin{equation}
\rho _{*}=\sum_{mm^{\prime }}\overline{\rho (\omega _0)}_{mm^{\prime
}}(\omega _0,mm^{\prime }|+\sum_{mm^{\prime }}\int_0^\infty d\omega 
\overline{\rho (\omega )}_{mm^{\prime }}(\omega ,mm^{\prime }|  \label{2.10}
\end{equation}
Therefore, in a weak sense we have: 
\begin{equation}
W\lim_{t\rightarrow \infty }\rho (t)=\rho _{*}  \label{2.11}
\end{equation}
Thus, any quantum state weakly goes to a linear combination of the energy
diagonal states $(\omega _0,mm^{\prime }|$ and $(\omega ,mm^{\prime }|$ (the
energy off-diagonal states $(\omega \omega _0,mm^{\prime }|$, $(\omega
_0\omega ^{\prime },mm^{\prime }|$ and $(\omega \omega ^{\prime },mm^{\prime
}|$ are not present in $\rho _{*}$). This is the case if we observe and
measure the system evolution with {\it any possible observable of space }$%
{\cal O}${\it .} Then, from the observational (or generalized
coarse-graining) point of view, we have decoherence of the energy levels
when $t\rightarrow \infty $, even that, from the strong limit
(fine-graining) point of view the off-diagonal terms never vanish, they just
oscillate, since we cannot directly use the Riemann-Lebesgue theorem in the
operator equation (\ref{2.7}).

Some observations are in order:

i.- The real existence of the two singular parts of $O$ and $\rho $ is
assured by the physic of the problem. The singular part of the observables
is just a necessary generalization of the singular part of the Hamiltonian,
which has a singular part $|\omega )$ (eq. (\ref{2.4})). The states must
also be singular objects since, intuitively, we realize that a continuous by
continuous matrix will decohere in a matrix with some kind of singularity in
the diagonal. The method is precisely designed to deal with this object.

ii.- From eq. (\ref{2.9}) we can again see that what we are doing is just a
generalized version of coarse graining, where a projector on the
''relevant'' part of the system is defined. The ''relevant'' part of the
states $(\rho |$ is in our case $(\rho |O)$ for all $O\in {\cal O}$, i.e.
the ''projection'' of $\rho $ on the class of observables of the form given
in eq. (\ref{2.5}). An ''irrelevant'' projection would be a $(\rho
|O^{\prime })$ where $O^{\prime }\notin {\cal O}$.

\subsection{Decoherence in the other ''momentum'' dynamical variables.}

Having established the decoherence in the energy levels we must consider the
decoherence in the other dynamical variables $O_i$, of the CSCO where we are
working. We will call these variables ''momentum variables''. For the sake
of simplicity we will consider, as in the previous section, that the spectra
of these dynamical variables are discrete. As the expression of $\rho _{*}$
given in eq. (\ref{2.10}) involves only the time independent components of $%
\rho (t)$, it is impossible that a different decoherence process would take
place to eliminate the off-diagonal terms in the remaining $N$ dynamical
variables. Therefore, the only thing to do is to find if there is a basis
where the off-diagonal components of $\rho (\omega _0)_{mm^{\prime }}$ and $%
\rho (\omega )_{mm^{\prime }}$ vanish at any time before the final state is
reached. This basis in fact exists, it is constant in time, and it will be
called the {\it final pointer basis.}

Let us consider the following change of basis 
\begin{equation}
|\omega _0,r\rangle =\sum_mU(\omega _0)_{mr}|\omega _0,m\rangle ,\qquad
|\omega ,r\rangle =\sum_mU(\omega )_{mr}|\omega ,m\rangle ,  \label{diag}
\end{equation}
where $r$ and $m$ are short notations for $r\doteq \{r_1,...,r_N\}$ and $%
m\doteq \{m_1,...,m_N\}$, and $\left[ U(x)^{-1}\right] _{mr}=\overline{U(x)}%
_{rm}$ ($x$ denotes either $\omega _0<0$ or $\omega \in {\Bbb R}^{+}$).

The new basis $\{|\omega _0,r\rangle ,|\omega ,r\rangle \}$ verifies the
generalized orthogonality conditions 
\[
\langle \omega _0,r|\omega _0,r^{\prime }\rangle =\delta _{rr^{\prime
}},\quad \langle \omega ,r|\omega ^{\prime },r^{\prime }\rangle =\delta
(\omega -\omega ^{\prime })\delta _{rr^{\prime }},\quad \langle \omega
_0,r|\omega ,r^{\prime }\rangle =\langle \omega ,r|\omega _0,r^{\prime
}\rangle =0. 
\]

It is easy to obtain the components of the states $\rho \in {\cal S}$ in the
new basis 
\begin{eqnarray*}
\rho (\omega _0)_{rr^{\prime }} &=&\sum_{mm^{\prime }}\left[ U(\omega
_0)^{-1}\right] _{rm}\rho (\omega _0)_{mm^{\prime }}\left[ U(\omega
_0)\right] _{m^{\prime }r^{\prime }}, \\
\rho (\omega )_{rr^{\prime }} &=&\sum_{mm^{\prime }}\left[ U(\omega
)^{-1}\right] _{rm}\rho (\omega )_{mm^{\prime }}\left[ U(\omega )\right]
_{m^{\prime }r^{\prime }}, \\
\rho (\omega ,\omega ^{\prime })_{rr^{\prime }} &=&\sum_{mm^{\prime }}\left[
U(\omega )^{-1}\right] _{rm}\rho (\omega ,\omega ^{\prime })_{mm^{\prime
}}\left[ U(\omega ^{\prime })\right] _{m^{\prime }r^{\prime }}, \\
\rho (\omega _0,\omega ^{\prime })_{rr^{\prime }} &=&\sum_{mm^{\prime
}}\left[ U(\omega _0)^{-1}\right] _{rm}\rho (\omega _0,\omega ^{\prime
})_{mm^{\prime }}\left[ U(\omega ^{\prime })\right] _{m^{\prime }r^{\prime
}}, \\
\rho (\omega ,\omega _0)_{rr^{\prime }} &=&\sum_{mm^{\prime }}\left[
U(\omega )^{-1}\right] _{rm}\rho (\omega ,\omega _0)_{mm^{\prime }}\left[
U(\omega _0)\right] _{m^{\prime }r^{\prime }},
\end{eqnarray*}
As $\overline{\rho (\omega _0)}_{mm^{\prime }}=\rho (\omega _0)_{m^{\prime
}m}$ and $\overline{\rho (\omega )}_{mm^{\prime }}=\rho (\omega )_{m^{\prime
}m}$, it is possible to choose $U(\omega _0)$ and $U(\omega )$ in such a way
that the off-diagonal parts of $\rho (\omega _0)_{rr^{\prime }}$ and $\rho
(\omega )_{rr^{\prime }}$ would vanish, i.e. 
\[
\rho (\omega _0)_{rr^{\prime }}=\rho _r(\omega _0)\,\delta _{rr^{\prime
}},\qquad \rho (\omega )_{rr^{\prime }}=\rho _r(\omega )\,\delta
_{rr^{\prime }}. 
\]
Therefore, there is a final pointer basis for the observables given by $%
\{|\omega _0,rr^{\prime })$, $|\omega ,rr^{\prime })$, $|\omega \omega
_0,rr^{\prime })$, $|\omega _0\omega ^{\prime },rr^{\prime })$, $|\omega
\omega ^{\prime },rr^{\prime })\}$ and defined as in eq. (\ref{2.5'}). The
corresponding final pointer basis for the states $\{(\omega _0,rr^{\prime }|$%
, $(\omega ,rr^{\prime }|$, $(\omega \omega _0,rr^{\prime }|$, $(\omega
_0\omega ^{\prime },rr^{\prime }|$, $(\omega \omega ^{\prime },rr^{\prime
}|\}$ diagonalizes the time independent part of $\rho (t)$ and therefore it
diagonalizes the final state $\rho _{*}$%
\begin{equation}
\rho _{*}=W\lim_{t\rightarrow \infty }\rho (t)=\sum_r\rho _r(\omega
_0)(\omega _0,rr|+\sum_r\int_0^\infty d\omega \rho _r(\omega )(\omega ,rr|.
\label{RO1}
\end{equation}

Now we can define the {\it final exact pointer observables} \cite{Zurek} 
\begin{equation}
P_i=\sum_rP_r^i(\omega _0)|\omega _0,r\rangle \langle \omega
_0,r|+\int_0^\infty d\omega \sum_rP_r^i(\omega )|\omega ,r\rangle \langle
\omega ,r|.  \label{RO2}
\end{equation}
As $H$ and $P_i$ are diagonal in the basis $\{|\omega _0,r\rangle $, $%
|\omega ,r\rangle \}$, the set $\{H,P_i,...P_N\}$ is precisely the complete
set of commuting observables (CSCO) related to this basis, where $\rho _{*}$
is diagonal in the corresponding co-basis for states. For simplicity we
define the operators $P_i$ such that $P_r^i(\omega _0)=P_r^i(\omega )=r_i$,
thus 
\begin{equation}
P_i|\omega _0,r\rangle =r_i|\omega _0,r\rangle ,\qquad P_i|\omega ,r\rangle
=r_i|\omega ,r\rangle .  \label{RO3}
\end{equation}
Therefore $\{|\omega _0,r\rangle $, $|\omega ,r\rangle \}$ is the final
observers' pointer basis where there is a perfect decoherence in the
corresponding state co-basis. Moreover the generalized states $(\omega
_0,rr| $ and $(\omega ,rr|$ are constants of the motion, and therefore these
exact pointer observables have a constant statistical entropy and will be
''at the top of the list'' of Zurek's ''predictability sieve'' \cite{Zurek}.
The final pointer basis is therefore defined by the dynamics of the model
and by the quantum state considered.

Therefore:

i.- Decoherence in the energy is produced by the time evolution when $%
t\rightarrow \infty $.

ii.- Decoherence in the other dynamical variables can be seen if we choose
an adequate basis, namely the final pointer basis.

Essentially we have given a partial answer, for this kind of models, to the
fundamental question of Gell-Mann and Hartle \cite{GyHUCS} (precisely only
an answer in the case when $t\rightarrow \infty )$: For each $H$ and each
initial state $\rho $ there is only one final pointer basis and therefore
only one ''quasi-classical domain or realm''. \cite{GyH} \footnote{%
But of course this unique consistent set depends of the chosen space of
observable ${\cal O}$ (see more in appendix III)${\cal .}$}.

Our main result is eq. (\ref{RO1}): {\it When }$t\rightarrow \infty $ {\it %
then }$\rho (t)\rightarrow \rho _{*}$ {\it and in this state the dynamical
variables }$H,P_1,...,P_N$ {\it are well defined. Therefore the eventual
conjugated variables to these momentum variables (namely: configuration
variables, if they exist) are completely undefined.}

In fact, calling ${\Bbb L}_i$ the generator of the displacements along the
eventual configuration variable conjugated to $P_i$, we have $({\Bbb L}%
_i\rho _{*}|O)=(\rho _{*}|{\Bbb L}_i^{\dagger }O)=(\rho _{*}|[P_i,O])=0$ for
all $O\in {\cal O}$ as it can be proved by direct computation using eqs. (%
\ref{2.5}),(\ref{2.5''}), (\ref{2.10}), and (\ref{RO2}). Then ${\Bbb L}%
_i\rho _{*}=0,$ and $\rho _{*}$\ is homogeneous in these configuration
variables.

\subsection{Decoherence characteristic decaying time, the permanent quantum
states case, and the role of the environment.}

From the preceding section we may have the feeling that the process of
decoherence must be found in all the physical systems. It is not so and
there are two reasons:

i.- Characteristic decaying times can be computed using analytic
continuation technics, as in paper \cite{CyLI}. E. g. in particular models
we can find the characteristic times for the system (e. g. an oscillator)
and the field (e. g. the environments or bath) as below eq. (56) of the last
quoted paper. If the maximal characteristic time $\gamma ^{-1}$ is very
large, even if theoretically the decoherence process will always take place,
it will be so slow that the system will behave as a quantum one for a very
long time. Then there will be no measurable decoherence.

ii.- It may also happen that more than one of the $\gamma $ would be zero.
Then, Hamiltonian $H,$ has more than one bound state, let us say $n$ (or
even part of its spectrum is discrete). Then the first term of the r. h. s.
of eq. (\ref{2.8}) must be changed to 
\begin{equation}
\sum_{ij}\rho _{ji}O_{ij}e^{i(\omega _i-\omega _j)t}=\sum_i\rho
_{ii}O_{ii}+\sum_{i\neq j}\rho _{ji}O_{ij}e^{i(\omega _i-\omega _j)t}
\label{3.4}
\end{equation}
where $i,j=1,...n,$ and as the second term of the r. h. s. does not vanish,
decoherence does not take place. This is the case of a theoretical atom, not
coupled to the electromagnetic field, where the electrons will remain for
ever in their exited states, and they will never decay. Then the atom never
goes to a decohered state. But if the atom is coupled to an electromagnetic
field (that usually it is called the ''environment'', as in appendix II)
there will be only one bound state, the second term of the r. h. s. of eq. (%
\ref{3.4}) will be absent, and decoherence will occur. In fact, in many
examples the role of the ''environment'' is just to introduce a continuous
spectrum to be coupled in such a way that only one bound state remains and
the decoherence is complete. In other cases fluctuations (or imperfections)
of continuous nature take the role of the continuous spectrum and produce
the average and make the off diagonal term disappear. This is the case of
the spin recombination experiment (\cite{Ballentine} page. 180) that takes
place in a single crystal interferometer.

iii.- More generally, using only observables from a subset $\Omega \in {\cal %
O}$ we may only involve some components of the state functional, e. g. those
constructed with the eigenvectors of $H$ that eventually expand the space $%
\Omega .$ Then if we only consider the observables of $\Omega $ it may be
that the components of the state related with these observables become
decohered, because their decoherence times are small, while the other
components remain undecohered, because they have a larger decoherence time.
Then we will have a system which is partially decohered and partially not
decohered, (which in fact is the case of the universe where there are both
classical and quantum phenomena).

\section{The classical statistical limit.}

\subsection{Expansion in sets of classical motions.}

In this section we will use the Wigner integrals that introduce an
isomorphism between quantum observables $O$ and states $\rho $ and their
classical analogues $O^W(q,p)$ and $\rho ^W(q,p)$ \cite{Wigner}: 
\begin{eqnarray}
O^W(q,p) &=&\int d\lambda \,\langle q-\frac \lambda 2|O|q+\frac \lambda 2%
\rangle \,\exp (\frac{i\lambda p}\hbar )  \nonumber \\
\rho ^W(q,p) &=&\frac 1{\pi \hbar }\int d\lambda \,(\rho ||q+\lambda \rangle
\langle q-\lambda |)\,\exp (\frac{2i\lambda p}\hbar ).  \label{Wig}
\end{eqnarray}

It is possible to prove that $\int dq\,dp\,\rho ^W(q,p)=(\rho |I)=1$, but $%
\rho ^W$ is not in general non negative. It is also possible to deduce that 
\begin{equation}
(\rho ^W|O^W)=\int dq\,dp\,\rho ^W(q,p)O^W(q,p)=(\rho |O),  \label{mean}
\end{equation}
and therefore to the mean value in the classical Liouville space it
corresponds the mean value in the quantum Liouville space. Moreover, calling 
$L$ the classical Liouville operator, and ${\Bbb L}$ the quantum
Liouville-Von Neumann operator, we have 
\begin{equation}
L\left[ \rho ^W(q,p)\right] =\left[ {\Bbb L}\rho \right] ^W(q,p)+O(\hbar ),
\label{ara1}
\end{equation}
where $L\,\rho ^W(q,p)=i\left\{ H^W(q,p),\rho ^W(q,p)\right\} _{PB}$ and 
\begin{equation}
({\Bbb L}\rho |O)=(\rho |[H,O]).  \label{ara2}
\end{equation}
Finally, if $O=O_1O_2$, where $O_1$ and $O_2$ are two quantum observables,
we have 
\begin{equation}
O^W(q,p)=O_1^W(q,p)O_2^W(q,p)+O(\hbar ).  \label{pro}
\end{equation}

We will prove that the distribution function $\rho _{*}^W(q,p)$, that
corresponds to the state functional $\rho _{*}$ via the Wigner integral is a
non negative function of the classical constants of the motion, in our case $%
H^W(q,p)$, $P_1^W(q,p)$,..., $P_N^W(q,p),$ obtained from the corresponding
quantum operators $H$, $P_1$,..., $P_N$.

From eq. (\ref{RO1}) we have: 
\begin{equation}
\rho _{*}=W\lim_{t\rightarrow \infty }\rho (t)=\sum_r\rho _r(\omega
_0)(\omega _0,rr|+\sum_r\int_0^\infty d\omega \rho _r(\omega )(\omega ,rr|,
\label{5.2}
\end{equation}
so we must compute: 
\begin{equation}
\rho _{\omega r}^W(q,p)\doteq \left( \frac 1{\pi \hbar }\right) ^{N+1}\int
(\omega ,rr||q+\lambda \rangle \langle q-\lambda |)e^{2ip\lambda }d\lambda
\label{5.3}
\end{equation}
We know from \cite{CyLII} section II C (or we can directly prove from eqs. (%
\ref{RO1}-\ref{RO3})) that 
\[
\]
\begin{equation}
\quad (\omega _0,rr|H^n)=\omega _0^n,\quad (\omega ,rr|H^n)=\omega ^n,\quad
(\omega _0,rr|P_i^n)=r_i^n,\quad (\omega ,rr|P_i^n)=r_i^n,  \label{5.4}
\end{equation}

for $i=1,...,N$ and $n=0,1,2,...$ Using the relation (\ref{pro}) between
quantum and classical products of observables and relation (\ref{mean})
between quantum and classical mean values, in the limit $\hbar \rightarrow 0$
(we will consider that we always take this limit when we refer to classical
equations below) we deduce that the characteristic property of the
distribution $\rho _{\omega r}^W(q,p)$, that corresponds to the state
functional $(\omega ,rr|$, is: 
\begin{equation}
\int \rho _{\omega r}^W(q,p)[H^W(q,p)]^ndqdp=\omega ^n,\quad \int \rho
_{\omega r}^W(q,p)[P_i^W(q,p)]^ndqdp=r_i^n,  \label{5.5}
\end{equation}
for any natural number $n.$ Thus $\rho _{\omega r}^W(q,p)$ must be the
functional \footnote{%
We have omitted the $O(\hbar )$ of eqs. (\ref{ara1}) and (\ref{pro}). If we
reintroduce these $O(\hbar )$ we will see that eqs. (\ref{5.6}) and (\ref
{5.6'}) are only valid in the limit $\hbar \rightarrow 0.$ If $\hbar $ is
only very small the $\delta $ are just functions strongly peaked at the zero
value of their variables.}: 
\begin{equation}
\rho _{\omega r}^W(q,p)=\delta (H^W(q,p)-\omega )\delta
(P_1^W(q,p)-r_1)...\delta (P_N^W(q,p)-r_N).  \label{5.6}
\end{equation}
For the distribution $\rho _{\omega _0r}^W(q,p)$, that corresponds to the
state functional $(\omega _0,rr|$, we obtain 
\begin{equation}
\rho _{\omega _0r}^W(q,p)=\delta (H^W(q,p)-\omega _0)\delta
(P_1^W(q,p)-r_1)...\delta (P_N^W(q,p)-r_N).  \label{5.6'}
\end{equation}
Therefore, going back to eq. (\ref{5.2}) and since the Wigner relation is
linear, we have: 
\begin{equation}
\rho _{*}^W(q,p)=\sum_r\rho _r(\omega _0)\rho _{\omega
_0r}^W(q,p)+\sum_r\int_0^\infty d\omega \rho _r(\omega )\rho _{\omega
r}^W(q,p).  \label{5.7}
\end{equation}
Also we obtain $\rho _{*}^W(q,p)\geq 0$, because $\rho _r(\omega _0)$ and $%
\rho _r(\omega )$ are non negative.

Therefore, the classical state $\rho _{*}^W(q,p)$ is a linear combination of
the generalized classical states $\rho _{xr}^W(q,p)$ (where $x$ is either $%
\omega _0$ or $\omega $), having well defined values $x$, $r_1$,..., $r_N$
of the classical observables $H^W(q,p)$, $P_1^W(q,p)$,..., $P_N^W(q,p)$ and
the corresponding classical canonically conjugated variables completely
undefined since the $\rho _{xr}^W(q,p)$ are not functions of these
variables. {\it So we reach, in the classical case, to the same conclusion
than in the quantum case (see end of subsection II B). }But now all the
classical canonically conjugated variables $a_0,a_1,...,a_N$ do exist since
they can be found solving the corresponding Poisson brackets differential
equations.

As the momenta $H^W,P_1^W,...,P_N^W$, or any function of these momenta, that
we will call generically $\Pi ,$ are also constant of the motion, then we
have $\frac d{dt}\Pi =-\partial H/\partial \alpha =0$, where $\alpha $ is
the classically conjugated variable to $\Pi .$ So $H$ is just a function of
the $\Pi $ and: 
\begin{equation}
\frac d{dt}\alpha =\frac{\partial H(\Pi )}{\partial \Pi }=\varpi (\Pi
)=const.  \label{5.9}
\end{equation}
so: 
\begin{equation}
\alpha _j(t)=\varpi _j(\Pi )t+\alpha _j(0),\qquad j=0,1,...,N.  \label{5.10}
\end{equation}
Thus (going back to the old coordinates) in the set of classical motions
contained in the densities (\ref{5.6}) and (\ref{5.6'}) the momenta $%
H,P_1,...,P_N$, are completely defined and the origin of the corresponding
motions, that we will respectively call$^{}$ $a_0(0)$, $a_1(0)$,...and $%
a_N(0)$, are completely undefined, in such a way that the motions
represented in the last equation homogeneously fill the surface where $H^W$, 
$P_1^W$,..., and $P_N^W$, have constant values, which now turns out to be a
usual torus of phase space \footnote{%
If $H^W$, $P_1^W$ ,...,$P_N^W$ are isolating constants of the motion and the
tori are not broken \cite{Tabor}.}. This is the interpretation that we give
to the densities (\ref{5.6}) and (\ref{5.6'}) which are just functions of
the variables $H^W$, $P_1^W$,..., $P_N^W,$ but they are not of the classical
conjugated variables $a_0$, $a_1$,..., $a_N$.

Then, eq. (\ref{5.7}) can be considered as the expansion of $\rho
_{*}^W(q,p) $ in the sets of classical motions contained in $\rho
_{xr}^W(q,p),$ each one with a probability $\rho _r(x)$ ($x=\omega _0,\omega 
$).

Summing up:

i.- We have shown that the quantum state functional $\rho (t)$ evolves to a
final diagonal state $\rho _{*}$.

ii.- This quantum state $\rho _{*}$ has $\rho _{*}^W(q,p)$ as its
corresponding classical density.

iii.- This classical density can be decomposed in sets of classical motions
where $H^W$, $P_1^W$,..., $P_N^W$ remain constant. The origin of these
motions: $a_0(0),a_1(0),...,a_N(0)$ are homogeneously distributed.

iv.- From eqs. (\ref{5.6}-\ref{5.7}) we obtained that $\rho _{*}^W(q,p)=$

$f(H^W(q,p),P_1^W(q,p),...,P_N^W(q,p))\geq 0$.

\subsection{Expansion in terms of classical motions.}

We can now expand the densities given in eqs. (\ref{5.6}-\ref{5.7}) in terms
of classical motions. In fact, since

\begin{equation}
\int \prod_{i=0}^N\delta (a_i(q,p)-a_i(t))\prod_{i=0}^Nda_i(0)=1
\label{5.12}
\end{equation}
where $a_j(t)=\varpi _j(P^W)t+a_j(0)$, we can write eq. (\ref{5.7}) as: 
\begin{eqnarray}
\rho _{*}^W(q,p) &=&\int \sum_r\rho _r(\omega _0)\rho _{\omega
_0r}^W(q,p)\prod_{i=0}^N\delta (a_i(q,p)-a_i(t))\prod_{i=0}^Nda_i(0)+ 
\nonumber \\
&&\int \sum_r\int_0^\infty d\omega \rho _r(\omega )\rho _{\omega
r}^W(q,p)\prod_{i=0}^N\delta (a_i(q,p)-a_i(t))\prod_{i=0}^Nda_i(0).
\label{5.13}
\end{eqnarray}
We define 
\begin{equation}
\rho _{x,r,a(0)}^W(q,p,t)\doteq \delta (H^W(q,p)-x)\delta
(P_1^W(q,p)-r_1)...\delta (P_N^W(q,p)-r_N)\delta (a_0(q,p)-a_0(t))...\delta
(a_N(q,p)-a_N(t)),  \label{5.14}
\end{equation}
which corresponds to the classical distribution of a motion with momenta $x$%
, $r_1$,..., $r_N$ and initial conditions $a_0(0)$,..., $a_N(0)$, and
therefore to a single classical motion. So we can write eq. (\ref{5.13}) as 
\begin{equation}
\rho _{*}^W(q,p)=\int \sum_r\rho _r(\omega _0)\rho _{\omega
_0,r,a(0)}^W(q,p,t)\prod_{i=0}^Nda_i(0)+\int \sum_r\int_0^\infty d\omega
\rho _r(\omega )\rho _{\omega ,r,a(0)}^W(q,p,t)\prod_{i=0}^Nda_i(0).
\label{5.15}
\end{equation}
So we have proved eq. (\ref{0.6}) as we promised in the introduction.

The densities $\rho _{x,r,a(0)}^W(q,p,t)$ represent a point in phase space
with momenta $H^W=x$, $P_1^W=r_1$,..., $P_N^W=r_N$ and coordinates $%
a_j(t)=\varpi _j(P^W)t+a_j(0)$, i.e. they represent single classical
trajectories.

Then we have obtained the final classical limit. When $t\rightarrow \infty $
the quantum state functional $\rho $ becomes a diagonal state $\rho _{*}.$
The corresponding classical distribution $\rho _{*}^W(q,p)$ can be expanded
as a linear combination of density functions $\rho _{\omega
_0,r,a(0)}^W(q,p,t)$ and $\rho _{\omega ,r,a(0)}^W(q,p,t)$, representing
classical trajectories, each one weighted by their corresponding
probabilities $\rho _r(\omega _0)$ and $\rho _r(\omega )$. As the limit when 
$t\rightarrow \infty $ of our quantum model we have obtained a statistical
classical mechanical model, \cite{Ballentine}, and the {\it classical
statistical realm }is obtained.

\section{Correlations and Localization.}

From many examples (e. g. \cite{Giulini}) we know that eventually
correlations and the localization appear when $t\rightarrow \infty ,$ at
least in some variables and in some quantum systems. E. g., in appendix B we
give an example obtained using our method, where we can see that
correlations appear in variables $Q$ and $P,$ when $t\rightarrow \infty $
(see eq. (\ref{X.16})). As this state with correlations is a final state let
as call it $\rho _{*}$ and let us see how it can be incorporated in our
formalism.

As $\rho _{*}$ is a final stationary state it can be decomposed as in eq. (%
\ref{RO1}). From eq. (\ref{2.5''}) we have: 
\begin{equation}
(\omega _0,rr|\omega _{0,}r^{\prime }r^{\prime })=\delta _{rr^{\prime
}},\quad (\omega ,rr|\omega _{,}r^{\prime }r^{\prime })=\delta _{rr^{\prime
}},\quad (\omega _0,rr|\omega _{,}r^{\prime }r^{\prime })=0  \label{4.1}
\end{equation}

Thus from eq. (\ref{4.1}) we have: 
\begin{equation}
(\rho _{*}|\omega _{0,}rr)=\rho _r(\omega _0),\quad (\rho _{*}|\omega
_{,}rr)=\rho _r(\omega )  \label{4.2}
\end{equation}
So, given $\rho _{*},$ endowed with correlations and computed by any method
(including ours, see appendix B) we can find the corresponding initial
conditions $\rho _r(\omega _0),\rho _r(\omega )$ that yield, when $%
t\rightarrow \infty $, to this correlated state\footnote{%
The remaining initial conditions: $\rho _{rr^{\prime }}(\omega _0),$ $\rho
_{rr^{\prime }}(\omega ),$ $\rho _{rr^{\prime }}(\omega ,\omega ^{\prime }),$
$\rho _{rr^{\prime }}(\omega _0,\omega ^{\prime }),$ $\rho _{rr^{\prime
}}(\omega ,\omega _0)$ are irrelevant since the corresponding terms
disappear when $t\rightarrow \infty .$}. In general, all final decohered
stationary states (but not any quantum state) can be decomposed in this way,
in particular our correlated state.

We can repeat all these formulae in the classical perspective of section III
using the relation (\ref{mean}) between quantum and classical symbols,
computing the initial conditions $\rho _r(\omega _0),\rho _r(\omega )$ using
classical formulas: 
\begin{equation}
(\rho _{*}^W(q,p)|\rho _{\omega _0r}^W(q,p))=\rho _r(\omega _0),\quad (\rho
_{*}^W(q,p)|\rho _{\omega r}^W(q,p))=\rho _r(\omega )  \label{4.3}
\end{equation}
In this way the correlation and localization phenomena can be incorporated
in our formalism. But it is difficult to use coordinates $x,r,a$ to directly
obtain the final state $\rho _{*}$ since this state looks quite unfamiliar
in these coordinates but it turns out to be the minimal uncertainty wave
packet if we study the problem in the usual coordinates $q,$ $p,$ as we
prove in the appendix B via an example\footnote{%
In fact, it is not possible to formulate a general theory in $q,$ $p$
coordinates because, in order to make the computations, we must know the
relation of these coordinates with the energy and other momenta, and this is
only defined in specific models.}. Furthermore the correlation phenomenon
only appears if the potential and the initial conditions are such that all
the trajectories with non negligible probability are concentrated by the
dynamics eventually yielding a ''maximally localized'' or ''minimal
uncertainty'' wave packet (as in the example of appendix B). It is difficult
to see this fact in the abstract unfamiliar frame of the coordinates $x,r,a$%
, because the potentials are hidden by the diagonalization even if the
initial conditions are obviously present (i. e,: in the choice $\rho
_r(\omega _0),$ $\rho _r(\omega )).$ Anyhow the phenomenon is there, since
we obtain localization when $t\rightarrow \infty .$ In this way we can
consider the localized wave packet like a single classical system and the
limit {\it classical statistical mechanics }$\rightarrow $ {\it classical
mechanics }is obtained because the motion of the wave packet satisfies the
classical equations (\ref{5.10}) as in all the trajectories. Now the
processes ''a'' and ''b'' are explained and the limit {\it statistical
quantum mechanics}$\rightarrow ${\it classical mechanics }is completed. The 
{\it classical realm} is appeared.

\section{Comments and conclusions.}

Some observation are in order:

\subsection{Sketch of the classical limit.}

Using the final pointer basis obtained in section IIB, the time dependant
Wigner function (namely the diagonalized version of eq. (\ref{2.7})) is 
\begin{eqnarray}
\rho ^W(q,p,t) &=&\sum_r\overline{\rho _r(\omega _0)}\rho _{\omega
_0r}^W(q,p)+\sum_r\int_0^\infty d\omega \overline{\rho _r(\omega )}\rho
_{\omega r}^W(q,p)+\sum_{rr^{\prime }}\int_0^\infty d\omega \overline{\rho
(\omega ,\omega _0)}_{rr^{\prime }}e^{i(\omega -\omega _0)t}\rho _{\omega
\omega _0rr^{\prime }}^W(q,p)+  \nonumber \\
&&\sum_{rr^{\prime }}\int_0^\infty d\omega ^{\prime }\overline{\rho (\omega
_0,\omega ^{\prime })}_{rr^{\prime }}e^{i(\omega _0-\omega ^{\prime })t}\rho
_{\omega _0\omega rr^{\prime }}^W(q,p)+\sum_{rr^{\prime }}\int_0^\infty
d\omega \int_0^\infty d\omega ^{\prime }\overline{\rho (\omega ,\omega
^{\prime })}_{rr^{\prime }}e^{i(\omega -\omega ^{\prime })t}\rho _{\omega
\omega ^{\prime }rr^{\prime }}^W(q,p)  \nonumber \\
&=&\rho _{*}^W(q,p)+\Delta \rho (q,p,t),  \label{6.2}
\end{eqnarray}
where the coefficients $\rho _r(\omega _0)$ and $\rho _r(\omega )$ are the
probabilities of each ''classical'' final history, and $\Delta \rho (q,p,t)$
corresponds to ''quantum'' non-decohered histories\footnote{%
We do not write these histories in detail since they will disappear in a
moment.}. It is clear that when $t\rightarrow \infty $ (really after a
decoherence time $\gamma ^{-1})$ the terms corresponding to these histories
vanish according to Riemann-Lebesgue theorem (see paper \cite{CyLII}).

Now we know:

i.- That when $\hbar \rightarrow 0,$ $\rho ^W(q,p,t)$ satisfies the
classical Liouville equation, namely the laws of classical mechanics.

ii.- That $\rho _{*}^W(q,p)\geq 0$, but that the second term $\Delta \rho
(q,p,t)$ is not positive definite, so $\rho ^W(q,p,t)=\rho
_{*}^W(q,p)+\Delta \rho (q,p,t)$ is not positive definite and therefore
cannot be considered as a classical density. Nevertheless $\Delta \rho $
vanishes when $t\rightarrow \infty $, so that $\rho ^W(q,p,t)$ is ''almost''
positive definite. Therefore with an adequate ''coarse-graining'' \footnote{%
Or, in our language, observed by an observer space smaller than ${\cal O}$.}%
, the averaged $\rho ^W(q,p,t)$ may be positive definite, for $t\gg \gamma
^{-1}$, and also would satisfy the classical Liouville equation, in its way
towards classical equilibrium. This $\rho ^W(q,p,t)$ would be a ''classical
limit'' before equilibrium. We will follow this line of research elsewhere.
For the moment it is clear that the non-final pointer basis has for limit
the final pointer basis when $t\rightarrow \infty .$ This fact may help us
to find both the non final pointer basis and the classical limit.

\subsection{Local vs. global equilibrium.}

In the last subsection we have considered the case where classicality is
reached before classical equilibrium. In this subsection we will see this
process in a different way. We can decompose the global system in a set of
local systems. In these subsystems, classical local equilibrium can be
reached after a time $\gamma ^{-1}$ with positive definite local equilibrium
densities (and, among other things we will be able to define a classical
local equilibrium entropy \cite{Mansur}). Since the classical local
equilibrium densities of each subsystem are positive definite the classical
density of the whole system will be positive definite. But the whole system
is out of equilibrium and its evolution is defined by the classical
Liouville equation. Now, as the system is already classical, since all its
parts are classical, we can study its evolution towards equilibrium with a
classical relaxation time $\tau ,$ that we are supposing $\tau >\gamma
^{-1}, $ with the usual classical methods (and the global entropy of
classical phenomenological thermodynamics will become maximal).

Of course, if the interaction is such that $\tau <\gamma ^{-1}$ this process
is not possible and we will directly reach to the final equilibrium without
a previous stage of local equilibrium. But the case $\tau >\gamma ^{-1}$ is
the usual one and it takes place if the global-long-range interactions are
unimportant with respect to the local-short-range interactions. Therefore
the system can be decomposed into a set of many weakly interacting
subsystems that can be considered as quasi-isolated. The local interactions
transform these quantum subsystems in classical subsystems in equilibrium,
and the system can be described as a non homogeneous distribution of local
subsystem with classical momenta $x,$ $l_1,...,l_N.$ These momenta, and the
subsystems density are different in each of them, producing a state of
classical non-equilibrium, that will reach equilibrium, due to the
global-long-range forces at a time $\tau >\gamma ^{-1}($see \cite{Halliwell}%
).

\subsection{Final Conclusion.}

Using the interplay of observables and states, that we have considered as
functionals over the space of observables, we have found an {\it exact final 
}pointer basis and an {\it intrinsically} consistent set of final histories.
So, given a Hamiltonian $H$ and a state $\rho $ we have found the exact
final pointer basis $\{|x,r_1,...,r_N\rangle \}$ and we have shown that $%
\rho _{*}^W$, the Wigner function of $\rho _{*}$, can be expanded in Wigner
functions corresponding to the co-basis $(x,r_1,...,r_N|$. To obtain this
results or similar ones almost all the authors use coarse graining methods
based in projectors and try to obtain a limit. So they essentially use the
weak limit of eq. (\ref{2.9}), namely: 
\[
\lim_{t\rightarrow \infty }(\rho (t)|O)=(\rho _{*}|O),\forall O\in {\cal O} 
\]
But, at least in the classical case, we know that this weak limit exists if
and only if the system is mixing \cite{Mackey}. And the system is mixing if
it has a continuous spectrum ( papers \cite{CyLII}, \cite{Antoniou}, \cite
{LyCE}) and the present paper can be considered as an extension of the
theorem that says that the mixing evolutions have a weak limit towards
equilibrium \cite{RS} but now formulated in the quantum case. Thus the only
way to deal with the problem (at least in the limit $t\rightarrow \infty )$
in an exact way is to use a method , as ours, specially adapted to deal with
the singularities inherent to that continuous spectrum. If not we are
condemned to only do approximate calculations.

Nevertheless approximated methods are important and, in some cases,
unavoidable to obtain the non-final pointer basis, but they can be better
understood if they are compared with exact methods. We will continue our
research following this line.

\begin{center}
{\bf ACKNOWLEDGMENTS.}
\end{center}

We are very grateful for the hospitality of Jonathan Halliwell, and the
Imperial College (London), where the research of this subject was began by
one of us (M.C.). This work was partially supported by grants CI1$^{*}$%
-CT94-0004 and PSS$^{*}-0992$ of the European Community, PID 3183/93 of
CONICET, EX053 of the Buenos Aires University, and also grants from
Fundaci\'{o}n Antorchas and OLAM Foundation.

\appendix 

\section{Comparison with the literature.}

In this appendix we would like to compare our method with those that can be
found in the literature, where the models are studied using the variables $Q$
and $P.$

Let us first see what is the shape of the diagonal states $(\omega _0|,$ $%
(\omega |$ or $\rho _{*}$ in $Q$ and $P.$ In order to determinate the
diagonal states, like $\rho _{*},$ in the configuration and momentum basis
(and also to find the correlations in these variables) we are forced to go
to a particular model were the relation among $H,Q,$and $P$ is defined$.$ We
consider a coupled system of an oscillator and a bath such that the
Hamiltonian reads \cite{CGG} : 
\begin{equation}
H=\frac 12\Omega (p^2+q^2)+\frac 12\int \omega (p_\omega ^2+q_\omega
^2)d\omega +\lambda \int V(\omega )(qq_\omega +pp_\omega )d\omega
\label{2.A}
\end{equation}
where the first term corresponds to the oscillator (with bound eigenstates $%
|\omega _i\rangle $ and ground state $|\omega _0\rangle $), the second term
to a field: the ''bath'' (with eigenstates $|\omega _1,\omega _2,...,\omega
_n\rangle $), and the third is an interaction term (with a $p\leftrightarrow
q$ symmetry) that leaves just one set of possible final states 
\begin{equation}
\rho =(\omega _0|=\sum_n\int \prod d\omega _id\omega _i^{\prime }\rho
_{0\omega _1,\omega _2,...,\omega _n\omega _1^{\prime },\omega _2^{\prime
},...,\omega _n^{\prime }}(\omega _{0,}\omega _1,\omega _2,...,\omega
_n\omega _1,\omega _2^{\prime },...,\omega _n^{\prime }|  \label{2.A'}
\end{equation}
namely, the oscillator in the ground state and the bath in any state (see 
\cite{CGG}). The $(\omega _{0,}\omega _1,\omega _2,...,\omega _n\omega
_1^{\prime },\omega _2^{\prime },...,\omega _n^{\prime }|$ are generated by
the dressed operators that we will define in eq. (\ref{X.5}).

Then let us try to get an idea of the form of $\rho _{*}$ via a heuristic
reasoning based on the symmetry of the Hamiltonian (\ref{2.A}). If we
consider a quantum state $\rho $ and the position operator $Q$ (that
symbolizes either the operator $q$ or any of the operators $q_\omega )$ and
the momentum operator $P$ (that symbolizes either $p$ or $p_\omega )$ in the
usual case we will have: 
\[
(\Delta Q)^2=Tr(Q^2\rho )-[Tr(Q\rho )]^2=\langle Q^2\rangle -\langle
Q\rangle ^2 
\]
\begin{equation}
(\Delta P)^2=Tr(P^2\rho )-[Tr(P\rho )]^2=\langle P^2\rangle -\langle
P\rangle ^2  \label{2.B}
\end{equation}
When $\rho $ is a functional, we can generalize these equation as: 
\[
(\Delta Q)^2=(\rho |Q^2)-[(\rho |Q)]^2=\langle Q^2\rangle -\langle Q\rangle
^2 
\]
\begin{equation}
(\Delta P)^2=(\rho |P^2)-[(\rho |P)]^2=\langle P^2\rangle -\langle P\rangle
^2  \label{2.B'}
\end{equation}
and in general $(\Delta Q)^2\neq (\Delta P)^2.$ But if $\rho $ is the
diagonal state $\rho _{*}$ of eq. (\ref{2.10}) (or the states $(\omega _0|,$ 
$(\omega |)$ we will have: 
\begin{equation}
(\rho |Q)=(\omega _0|Q),\text{ }(\rho |Q^2)=(\omega _0|Q^2),\text{ }(\rho
|P)=(\omega _0|P),\text{ }(\rho |P^2)=(\omega _0|P^2)  \label{2,C}
\end{equation}
So, as $H$ has a $q-p-$symmetry everything is symmetric under the
transformation $p\leftrightarrow q$ (or $q\rightarrow -i\frac \partial {%
\partial q},-i\frac \partial {\partial q}\rightarrow q)$ and therefore $%
(\Delta Q)^2=(\Delta P)^2.$ This would not be the case if $\rho $ would not
be diagonal in the basis where $H$ is diagonal, since $Q$ and $P$ are not
diagonal in this basis, e. g. $\rho $ could commute with $P$ but not with $Q$
showing, in this case, a clear asymmetry $P\leftrightarrow Q.$ Thus our
diagonal states are states such that $\Delta Q=\Delta P$. Namely: for the
ground state we have $\Delta p=\Delta q,$ really a well known fact. If now
we introduce in (\ref{2.A}) a small asymmetric interaction $\lambda ^{\prime
}W$ $(\lambda ^{\prime }\ll 1)$ we will have $\Delta Q\cong \Delta P$. On
the contrary if the interaction is $\lambda ^{\prime }W(q,q_\omega )$ ($%
\lambda ^{\prime }\gg 1)$ the Hamiltonian $H$ can be neglected and the
diagonal states will be position eigenvalues (so our results coincide with
those of refs. \cite{Zurek} and \cite{Juanpa}, see a detailed example below).

\section{An example of correlations and localization.}

There are systems, e. g. the one of appendix A, with variables $Q$ and $P$
and a bath, where the interaction is such that $Q$ and $P$ become
correlated. Namely the evolution makes both $\Delta Q$ and $\Delta P$
bounded, and a wave packet appears that eventually becomes a minimal
uncertainty wave packet, when $t\rightarrow \infty ,$ then maximal
localization appears in the usual way as promised in section IV. As an
example, let us now find the correlations between $Q$ and $P$ in the model
of Hamiltonian (\ref{2.A}) using our method as explained above and in the
paper \cite{LyCE}. Let us first write the Hamiltonian (\ref{2.A}) using
creation and annihilation operators.

\begin{equation}
H=\Omega b^{\dagger }b+\int d{\bf k}\omega _ka_{{\bf k}}^{\dagger }a_{{\bf k}%
}+\int d{\bf k}V_k(a_{{\bf k}}^{\dagger }b+b^{\dagger }a_{{\bf k}}),\qquad
\omega _k=k,\qquad k=\left| {\bf k}\right|  \label{X.1}
\end{equation}
The coordinate $q$ and the momentum $p$ of the oscillator can be expressed
as a function of the $b^{\dagger }$ and $b$ as: 
\begin{equation}
q=\left( \frac \hbar {2m\Omega }\right) ^{\frac 12}(b^{\dagger }+b);\qquad
p=i\left( \frac{m\hbar \Omega }2\right) ^{\frac 12}(b^{\dagger }-b).
\label{X.2}
\end{equation}

We can adimensionalize the last equation defining $Q$ and $P$ such that: 
\begin{equation}
q=\left( \frac \hbar {m\Omega }\right) ^{\frac 12}Q,\qquad p=\left( m\hbar
\Omega \right) ^{\frac 12}P  \label{X.3}
\end{equation}
Then: 
\begin{equation}
Q=\frac 1{\sqrt{2}}(b^{\dagger }+b),\qquad P=i\frac 1{\sqrt{2}}(b^{\dagger
}-b)  \label{X.3'}
\end{equation}
\begin{equation}
b=\frac 1{\sqrt{2}}(Q+iP),\qquad b^{\dagger }=\frac 1{\sqrt{2}}(Q-iP)
\label{X.3"}
\end{equation}

In the Heisenberg representation the operator $Q$ evolves as: 
\begin{equation}
Q(t)=\frac 1{\sqrt{2}}(b^{\dagger }(t)+b(t))=\frac 1{\sqrt{2}}\int d{\bf k}%
V_k\left( \frac 1{\eta _{-}(k)}A_{{\bf k}}^{\dagger }e^{i\omega _kt}+\frac 1{%
\eta _{+}(k)}A_{{\bf k}}e^{-i\omega _kt}\right) ,  \label{X.4}
\end{equation}
where: 
\begin{equation}
A_{{\bf k}}^{\dagger }=a_{{\bf k}}^{\dagger }+\frac{V_k}{\eta _{+}(k)}\left(
b^{\dagger }+\int \frac{d{\bf k}^{\prime }V_{k^{\prime }}a_{{\bf k}^{\prime
}}^{\dagger }}{\omega _k-\omega _{k^{\prime }}+i0}\right) ,\qquad A_{{\bf k}%
}=a_{{\bf k}}+\frac{V_k}{\eta _{-}(k)}\left( b+\int \frac{d{\bf k}^{\prime
}V_{k^{\prime }}a_{{\bf k}^{\prime }}}{\omega _k-\omega _{k^{\prime }}+i0}%
\right) .  \label{X.5}
\end{equation}
The functions $\eta _{-}(k)$ and $\eta _{+}(k)$ and all the details of the
calculations can be found in paper \cite{LyCE}.

Let us consider the initial conditions 
\begin{equation}
\langle Q\rangle _{t=0}=Q_0,\qquad \langle P\rangle _{t=0}=P_0  \label{X.6}
\end{equation}
for the oscillator, and also 
\begin{equation}
\langle a_{{\bf k}}^{\dagger }\rangle _{t=0}=\langle a_{{\bf k}}\rangle
_{t=0}=0,  \label{X.6'}
\end{equation}
which corresponds to the field being initially in its ground state.
Therefore: 
\begin{equation}
\langle b\rangle _{t=0}=\frac 1{\sqrt{2}}(Q_0+iP_0),\qquad \langle
b^{\dagger }\rangle _{t=0}=\frac 1{\sqrt{2}}(Q_0-iP_0)  \label{X.7}
\end{equation}
and for the time evolution of the mean value of the coordinate and momentum
of the oscillator we obtain 
\begin{equation}
\langle Q\rangle _t=\frac 1{\sqrt{2}}\int d{\bf k}\frac{V_k^2}{\eta
_{-}(k)\eta _{+}(k)}\left( e^{i\omega _kt}\langle b^{\dagger }\rangle
_{t=0}+e^{-i\omega _kt}\langle b\rangle _{t=0}\right) ,  \label{X.8}
\end{equation}
\begin{equation}
\langle P\rangle _t=\frac i{\sqrt{2}}\int d{\bf k}\frac{V_k^2}{\eta
_{-}(k)\eta _{+}(k)}\left( e^{i\omega _kt}\langle b^{\dagger }\rangle
_{t=0}-e^{-i\omega _kt}\langle b\rangle _{t=0}\right) .  \label{X.9}
\end{equation}

The oscillating time dependent factors inside the integrals produce the
vanishing of both $\langle Q\rangle _t$ and $\langle P\rangle _t$ for very
long times. We can study the poles of the analytic extension of the factor $%
\frac{V_k^2}{\eta _{-}(k)\eta _{+}(k)}$, as in paper \cite{CyLI} and prove
that the trajectory of $\langle Q\rangle _t$ and $\langle P\rangle _t$ in
the phase space of the oscillator is a spiral ending at $\langle Q\rangle
=\langle P\rangle =0$.

Now we would like to compute $\Delta q$ and $\Delta p$ as a function of
time. In addition to equations (\ref{X.6}) and (\ref{X.6'}) let us assume
the following initial conditions for the oscillator 
\begin{equation}
\langle bb^{\dagger }\rangle _{t=0}=\beta ,\qquad \langle bb\rangle
_{t=0}=\alpha ,\qquad \langle b^{\dagger }b\rangle _{t=0}=1-\beta ,\qquad
\langle b^{\dagger }b^{\dagger }\rangle _{t=0}=\alpha ^{*},  \label{X.10'}
\end{equation}
being $\alpha $ and $\beta $ some arbitrary constants. If the field is in
its ground state, we also have 
\begin{equation}
\langle a_{{\bf k}}a_{{\bf k}^{\prime }}^{\dagger }\rangle _{t=0}=\delta ^3(%
{\bf k}-{\bf k}^{\prime }),\qquad \langle a_{{\bf k}}^{\dagger }a_{{\bf k}%
^{\prime }}\rangle _{t=0}=\langle a_{{\bf k}}^{\dagger }a_{{\bf k}^{\prime
}}^{\dagger }\rangle _{t=0}=\langle a_{{\bf k}}a_{{\bf k}^{\prime }}\rangle
_{t=0}=0.  \label{X.11}
\end{equation}
All other initial mean values of products of pairs of creation or
annihilation operators are zero. This means that we have taken the
oscillator in an arbitrary state and the field in the ground state as
initial conditions.

Therefore we have 
\begin{eqnarray}
\langle A_{{\bf k}}^{\dagger }A_{{\bf k}^{\prime }}^{\dagger }\rangle _{t=0}
&=&\frac{V_kV_{k^{\prime }}}{\eta _{+}(k)\eta _{+}(k^{\prime })}\alpha ^{*},
\nonumber \\
\langle A_{{\bf k}}^{\dagger }A_{{\bf k}^{\prime }}\rangle _{t=0} &=&\frac{%
V_kV_{k^{\prime }}}{\eta _{+}(k)\eta _{-}(k^{\prime })}(\beta -1),  \nonumber
\\
\langle A_{{\bf k}}A_{{\bf k}^{\prime }}\rangle _{t=0} &=&\frac{%
V_kV_{k^{\prime }}}{\eta _{-}(k)\eta _{-}(k^{\prime })}\alpha ,  \label{X.13}
\\
\langle A_{{\bf k}}A_{{\bf k}^{\prime }}^{\dagger }\rangle _{t=0} &=&\delta
^3({\bf k}-{\bf k}^{\prime })+\frac{V_{k^{\prime }}V_k}{\eta _{+}(k^{\prime
})(\omega _{k^{\prime }}-\omega _k+i0)}+\frac{V_kV_{k^{\prime }}}{\eta
_{-}(k)(\omega _k-\omega _{k^{\prime }}-i0)}+  \nonumber \\
&&+\frac{V_kV_{k^{\prime }}}{\eta _{-}(k)\eta _{+}(k^{\prime })}\beta +\frac{%
V_kV_{k^{\prime }}}{\eta _{-}(k)\eta _{+}(k^{\prime })}\int \frac{d{\bf k}%
^{\prime \prime }V_{k^{\prime \prime }}^2}{(\omega _k-\omega _{k^{\prime
\prime }}-i0)(\omega _{k^{\prime }}-\omega _{k^{\prime \prime }}+i0)}. 
\nonumber
\end{eqnarray}

The time evolution of the mean value of $Q(t)^2$ is given by 
\begin{eqnarray}
\langle Q(t)^2\rangle &=&\frac 12\langle \int d{\bf k}V_k\left( \frac 1{\eta
_{-}(k)}A_{{\bf k}}^{\dagger }e^{i\omega _kt}+\frac 1{\eta _{+}(k)}A_{{\bf k}%
}e^{-i\omega _kt}\right) \times  \nonumber \\
&&\times \int d{\bf k}^{\prime }V_{k^{\prime }}\left( \frac 1{\eta
_{-}(k^{\prime })}A_{{\bf k}^{\prime }}^{\dagger }e^{i\omega _{k^{\prime
}}t}+\frac 1{\eta _{+}(k^{\prime })}A_{{\bf k}^{\prime }}e^{-i\omega
_{k^{\prime }}t}\right) \rangle  \label{X.12}
\end{eqnarray}
Then, replacing equations (\ref{X.13}) in equation (\ref{X.12}) and always
using the Riemann-Lebesgue theorem we have: 
\begin{equation}
\lim_{t\rightarrow \infty }\langle Q(t)^2\rangle =\frac 12\int d{\bf k}\frac{%
V_k^2}{\eta _{-}(k)\eta _{+}(k)}=\frac 12,  \label{X.15}
\end{equation}
and therefore 
\[
\lim_{t\rightarrow \infty }\left[ \Delta Q(t)\right] ^2=\lim_{t\rightarrow
\infty }\left\langle (Q(t)-\langle Q(t)\rangle )^2\right\rangle =\frac 12. 
\]
Making an analogous calculation for $P$, we have: 
\begin{equation}
\lim_{t\rightarrow \infty }\Delta Q=\lim_{t\rightarrow \infty }\Delta P=%
\frac 1{\sqrt{2}}  \label{X.16}
\end{equation}
so the wave packet around the spiral trajectory evolves to a minimal
uncertainty symmetrical wave packet, showing the localization process in the
usual way. This proves the presence of correlations in our model.
Reestablishing the units, when $t\rightarrow \infty ,$ and introducing the
velocity $v$ we have: 
\begin{equation}
\Delta q=\left( \frac{\hbar ^2}{2m\Omega }\right) ^{\frac 12},\qquad \Delta
v=\left( \frac{\hbar ^2\Omega }{2m}\right) ^{\frac 12}  \label{X.17}
\end{equation}
This fact shows that the wave packet is more peaked for big a $m$ than for
small a $m.$ Then, in some models big mass particles can be considered as
classical while other remain quantum. Moreover if $\hbar \rightarrow 0$ the
uncertainties disappear. Also the classical limit of the oscillator has a
spiral motion in phase space. In fact, if using eq. (\ref{Wig}) we compute
the Wigner function corresponding to the matrix density, we will find the
motion of this classical density that will be centered in the spiral
trajectory and having, when $t\rightarrow \infty ,$ a symmetrical circle of
diameter ($\frac 12\hbar )^{\frac 12}$ as support \footnote{%
The cosmological models of papers \cite{Cosmo} are other examples that we
will further develop elsewhere.}.

\section{Decoherence of histories.}

From the section VA we can conclude that our notion of history of the system
is essentially contained in the state $\rho (t).$ This history ends in the
final equilibrium state $\rho _{*}$. In this appendix we will study the
relation of this notion with the usual histories formalism \cite{Halliwell}
and compare the results. The computation will turn out to be very simple for
two reasons:

i.- As in all the paper we will work only in the limit $t\rightarrow \infty $%
, where the existence of an exact final pointer basis will make all the
computations quite trivial.

ii.- Also we will only consider one space of observables ${\cal O}$ and
therefore just one set of final consistent histories \cite{Dowker}. The case
of many sets will be considered elsewhere.

Nevertheless we think that the results are of some interest since:

i.- For times $t\gg \gamma ^{-1}$ all the exact result obtained in the limit 
$t\rightarrow \infty $ can be considered as good approximations.

ii.-The existence of a space of observables ${\cal O,}$ where we can use the
Riemann-Lebesgue theorem, perhaps can be considered as a selection principle
to choose the physically relevant consistent set \cite{Kent}.

So let us begin giving the main definitions. Let us consider a time
depending basis of ${\cal H:}$ $\{|\alpha (t)\rangle \},$ and the
projectors: 
\begin{equation}
P_\alpha (t)=|\alpha (t)\rangle \langle \alpha (t)|  \label{C.1}
\end{equation}
such that they represent exhaustive and exclusive alternatives: 
\begin{equation}
\sum_\alpha P_\alpha =1,\text{ }P_\alpha P_\beta =\delta _{\alpha \beta
}P_\alpha  \label{C.2}
\end{equation}
We will call (fine-grained) {\it \ history }$\overrightarrow{\alpha }$ to a
string of time depending projectors \footnote{%
We can consider a more general case were the exclusive and exhausting set of
histories is different at every time $t_i$ and therefore the projectors are $%
P_{\alpha _i}^i(t_i).$ But this is not the usual case.}: 
\begin{equation}
C_{\overrightarrow{\alpha }}=P_{\alpha _1}(t_1)...P_{\alpha _n}(t_n),\text{ }%
t_1<...<t_n  \label{C.3}
\end{equation}

For a state $\rho $ we will call {\it decoherence matrix }to: 
\begin{equation}
M(\overrightarrow{\alpha }{\bf ,}\overrightarrow{\alpha }^{\prime })=C_{%
\overrightarrow{\alpha }}^{\dagger }\rho C_{\overrightarrow{\alpha }^{\prime
}}=P_{\alpha _n}(t_n)...P_{\alpha _1}(t_1)\rho P_{\alpha _1^{\prime
}}(t_1^{\prime })...P_{\alpha _n^{\prime }}(t_n^{\prime })  \label{C.4}
\end{equation}
We introduce this matrix because we consider it as the natural
generalization of the usual density matrix to the case where single
projectors are changed by histories.

We will call {\it decoherence functional }to: 
\begin{equation}
D(\overrightarrow{\alpha }{\bf ,}\overrightarrow{\alpha }^{\prime })=TrM(%
\overrightarrow{\alpha }{\bf ,}\overrightarrow{\alpha }^{\prime })
\label{C.5}
\end{equation}
that would be the generalization of the trace of an ordinary matrix 
\footnote{%
If some of the $\alpha $ are continuous indices, for them we must use the
generalization of the trace introduced in paper \cite{CyLII}}.

We will call {\it candidate probability for the history }$\overrightarrow{%
\alpha }$ to: 
\begin{equation}
p(\overrightarrow{\alpha })=TrM(\overrightarrow{\alpha }{\bf ,}%
\overrightarrow{\alpha })  \label{C.6}
\end{equation}

that would be the generalization of the usual probability. It is only a
''candidate probability'' because, at this stage, it does not satisfy the
axioms of the usual boolean probability theory.

If 
\begin{equation}
\mathop{\rm Re}%
D(\overrightarrow{\alpha }{\bf ,}\overrightarrow{\alpha }^{\prime })=0
\label{C.7}
\end{equation}
for $\overrightarrow{\alpha }{\bf \neq }\overrightarrow{\alpha }^{\prime } $
we will say that the set of histories is {\it consistent or weakly decoherent%
}, in this case it is proved that the set can be in principle submitted to
the ordinary boolean logic \cite{Omnes}, and the candidate probability can
be considered as the probability of each history.

If 
\begin{equation}
D(\overrightarrow{\alpha }{\bf ,}\overrightarrow{\alpha }^{\prime })=0
\label{C.8}
\end{equation}
for $\overrightarrow{\alpha }{\bf \neq }\overrightarrow{\alpha }^{\prime } $
we will say that the set has {\it medium decoherence}. Theorems about
records can be proved if the set of histories has this type of decoherence 
\cite{Halli}.

If 
\begin{equation}
M(\overrightarrow{\alpha }{\bf ,}\overrightarrow{\alpha }^{\prime })=0
\label{C.9}
\end{equation}
for $\overrightarrow{\alpha }{\bf \neq }\overrightarrow{\alpha }^{\prime } $
we will say that the set is {\it intrinsically consistent }\cite{Zurek} or
that it has {\it matrix decoherence}. Of course matrix decoherence implies
medium decoherence, and medium decoherence implies weak decoherence.

Let us now compare all these concepts with our formalism.

We choose: 
\begin{equation}
|\alpha (t_1)\rangle =|x,r_1,...,r_N;t_1\rangle  \label{C.10}
\end{equation}
where we have used the shorthand notation introduced above. The set of
operators $P_{\alpha (t)}=|\alpha (t)\rangle \langle \alpha
(t)|=|x,r_1,...,r_N;t\rangle \langle x,r_1,...,r_N;t|$ will be our ''final
pool of operators'' if we use the language of paper \cite{GyHUCS}. The
evolution of these operators will be:) 
\[
P_\alpha (t)=e^{-iH(t-t_1)}P_\alpha
(t_1)e^{iH(t-t_1)}=e^{-ix(t-t_1)}P_\alpha (t_1)e^{ix(t-t_1)}= 
\]
\begin{equation}
P_\alpha (t_1)=P_\alpha =|\alpha (0)\rangle \langle \alpha (0)|  \label{C-11}
\end{equation}
i. e.: these operators are constant. Then the projectors are time constant
and: 
\begin{equation}
C_{\overrightarrow{\alpha }}=P_\alpha  \label{C.12}
\end{equation}
and these histories can be labeled with the ordinary $\alpha $ instead of
the $\overrightarrow{\alpha }$ with the arrow.

In more detail let us first study our ''pool'' of projectors to compute eq. (%
\ref{C.4}) in our formalism and when $t\rightarrow \infty ,$ 
\begin{equation}
P_{\alpha (t)}=P_\alpha =|x,r_1,...,r_N\rangle \langle
x,r_1,...,r_N|=|x,r_1,...,r_N)  \label{D.1}
\end{equation}

1.-$r_1,...,r_N$ are discrete indices and the final stationary state $\rho
_{*}$ is diagonal in these indices, so this part of the problem is trivial.

2.-$x$ symbolizes $(\omega _0,\omega )$ where only $\omega $ is continuous,
so the treatment of $\omega _0$ is also trivial.

The problem is only $\omega $ so, for simplicity, let us only consider this
index. The projector reads: 
\begin{equation}
P_\omega =|\omega \rangle \langle \omega |=|\omega )  \label{D.2}
\end{equation}
So let us compute: 
\begin{equation}
P_\omega \rho _{*}P_{\omega ^{\prime }}=|\omega \rangle \langle \omega |\rho
_{*}|\omega ^{\prime }\rangle \langle \omega ^{\prime }|  \label{D.3}
\end{equation}
but first we must find the meaning of this symbol. In the discrete case we
have: 
\begin{equation}
|a\rangle \langle b|\rho |c\rangle \langle d|=|a\rangle Tr(\rho |c\rangle
\langle b|)\langle d|  \label{D.4}
\end{equation}
that can be generalized to the continuous case as:

\begin{equation}
|a\rangle \langle b|\rho |c\rangle \langle d|=|a\rangle (\rho ||c\rangle
\langle b|)\langle d|  \label{D.5}
\end{equation}
Thus: 
\begin{equation}
P_\omega \rho _{*}P_{\omega ^{\prime }}=||\omega \rangle \langle \omega
|\left[ \int \rho _{\omega "}(\omega "|d\omega "\right] |\omega ^{\prime
}\rangle \langle \omega ^{\prime }|=|\omega \rangle \left[ \int \rho
_{\omega "}(\omega "|\omega ^{\prime },\omega )d\omega "\right] \langle
\omega ^{\prime }|  \label{D.6}
\end{equation}
So, from eqs. (\ref{2.5''}) we have:

1.- If $\omega \neq \omega ^{\prime }$ it is $P_\omega \rho _{*}P_{\omega
^{\prime }}=0.$

2.- If $\omega =\omega ^{\prime }$ it is: 
\begin{equation}
P_\omega \rho _{*}P_{\omega ^{\prime }}=|\omega \rangle \left[ \int \rho
_{\omega "}(\omega "|\omega )d\omega "\right] \langle \omega |=|\omega
\rangle \left[ \int \rho _{\omega "}\delta (\omega "-\omega )d\omega
"\right] \langle \omega |=\rho _\omega |\omega \rangle \langle \omega |
\label{D.7}
\end{equation}
So with a symbolic obvious notation (that we will use from now on) we can
say that: 
\begin{equation}
P_\omega \rho _{*}P_{\omega ^{\prime }}=|\omega \rangle \rho _\omega \delta
_{\omega \omega ^{\prime }}\langle \omega ^{\prime }|  \label{D.8}
\end{equation}
If now we repeat the reasoning including all the trivial discrete indices we
will obtain the same result since $\rho _{*}$ is diagonal in these indices.
Then, when $t\rightarrow \infty $ we have that 
\begin{equation}
M(\overrightarrow{\alpha }{\bf ,}\overrightarrow{\alpha }^{\prime
})\rightarrow \delta _{\alpha \alpha ^{\prime }}\rho _\alpha |\alpha \rangle
\langle \alpha |  \label{C.16}
\end{equation}
and therefore we have final matrix decoherence in a time long enough. Then
we have found the final ''statistical classical domain or realm'' of
Gell-Mann and Hartle. In this way final classical behavior emerges from
quantum behavior and transition 2a'' appears in the histories formalism.
Essentially we have used the weak limit of eq. (\ref{2.11}) and the fact
that it is the only possible limit we can use, since $\rho $ is a functional
over the space ${\cal O.}$ But the choice of eq. (\ref{C.10}) has an extra
bonus: it decomposes the density matrix just in the way that was announced
in the introduction.

From the matrix decoherence we have medium decoherence and weak decoherence,
so we have proved that any quantum system, fulfilling the conditions
required in section II, has a set of final intrinsically consistent
histories, the essential conditions being the continuous spectrum and the
existence of just one ground state. Classically these histories will be the $%
\rho _{xr}^W(q,p)$ of eq.(\ref{6.2}). This exact final decoherence has being
obtained using the basis \{$|xr\rangle \}$, other near bases obviously yield
final approximate decoherence. Also basis \{$|xr\rangle \}$ will give
approximate decoherence in a time long enough.

But we must observe that in all cases where $P_\alpha (t)=P_\alpha =const.$
(even if eq. (\ref{C.10}) is not satisfy) we can immediately prove medium
decoherence with no reference to matrix decoherence. In fact, if $P_\beta
=|\beta \rangle \langle \beta |=const.$ we have: 
\begin{equation}
D(\overrightarrow{\beta }{\bf ,}\overrightarrow{\beta }^{\prime })=D(\beta
,\beta ^{\prime })=Tr(|\beta \rangle \langle \beta |\rho |\beta ^{\prime
}\rangle \langle \beta ^{\prime }|)=\langle \beta |\beta ^{\prime }\rangle
\langle \beta |\rho |\beta ^{\prime }\rangle =\delta _{\beta \beta ^{\prime
}}p(\beta )  \label{C.16'}
\end{equation}
These would be the case with the $P_\alpha $ of this section and also for
any constant $P_\beta $. This result seems very trivial but it is not. The
essential property of projector (\ref{D.1}) is that it is time constant, but
our formalism contains other time-constant projectors. If we go back to
section IIA we find: 
\begin{equation}
P_{\beta (t)}=P_\beta =|x,m_1,...,m_N\rangle \langle
x,m_1,...,m_N|=|x,m_1,...,m_N)  \label{E.1}
\end{equation}
namely the projectors related with the basis $\{x,m\}$ before the
diagonalization (\ref{diag}) that yields the basis $\{x,r\}.$ The $P_\beta $
are also time constants and yield medium decoherence (only the $P_a$ yield
matrix decoherence). The main fact is that in order to reach the classical
statistical mechanics of section III we must use the basis $\{x,r\}$ that
diagonalize $\rho _{*}$ in {\it all indices (}see eqs. (\ref{5.2}) to (\ref
{5.7})). Thus, since our demonstration is based in the matrix decoherence in
the basis $\{x,r\},$ these objects are essential for us. Only after this
demonstration we can speak of classical constants of the motion and
classical trajectories because only then we can pass from the quantum
formulae to the classical ones.

Then the last result can be translated as follows:

1.- There is final matrix decoherence between any pair of different sets of
constants ($x,r)$ i. e. between any pair of sets of classical trajectories
in the phase space. This set of sets of trajectories is intrinsically
consistent (see eq. (\ref{RO1})).

2.- But, of course, any set of functions of the $"r"$, such as the $"m"$,
will define equally well the set of classical trajectories. But the $"m"$ do
not provide a basis with good defined probabilities, as the $"r"$ does,
since in the basis $"m"$ the $\rho _{*}$ is not diagonal (see eq. (\ref{2.10}%
)). In this case the set of histories is consistent but not intrinsically
consistent.

So our point of view is that, even if all sets endowed of medium decoherence
can be considered as consistent sets, there is only one with physical
importance, the one with matrix decoherence, the only one which is an
''intrinsically consistent set''. This idea may help to find the selection
principle searched in papers \cite{Kent}.

Finally, if the potential and the initial conditions are such to privilege a
history (as in appendix B) the locations process ''b'' will take place and
we will have a unique classical object with a unique history $P_\alpha .$
Then we would find the final ''classical domain or realm'' of Gell-Mann and
Hartle.

We will end this section showing how several requirements necessary for a
efficient histories decoherence are satisfied by our formalism:

\subsection{Griffiths-Omn\`{e}s condition}

The Griffiths-Omn\`{e}s condition for consistency \cite{Omnes}, \cite
{Griffiths} is automatically satisfied since: 
\begin{equation}
\mathop{\rm Re}%
Tr[|\alpha \rangle \langle \alpha |\rho (1-|\alpha \rangle \langle \alpha
|)|\alpha \rangle \langle \alpha |]=0  \label{C.12'}
\end{equation}

\subsection{Permanence of the past.}

If we take our projectors from the pool of the projectors $|\alpha \rangle
\langle \alpha |$ the condition of permanence of the past \cite{GyHUCS} is
trivially satisfied , since a chain with a certain number of $|\alpha
\rangle \langle \alpha |$ can only be continued repeating this projector.
This is the most important property required in papers \cite{Kent}.

\subsection{Insensitivity.}

While quantum states are modified by the measurement processes, classical
states are not sensible to these measurements. This property of classical
states is called insensitivity \cite{Zurek}. The projector $P_{\alpha _i}$ =$%
|\alpha _i\rangle \langle \alpha _i|$ can be considered as a measurement
operator, so if $\rho _{before}$ is the state before the measurement and $%
\rho _{after\text{ }}$is the state after the measurement, we will have: 
\begin{equation}
\rho _{after}=\sum_iP_{\alpha _i}\rho _{before}P_{\alpha _i}=\sum_i|\alpha
_i\rangle \langle \alpha _i|\rho _{before}|\alpha _i\rangle \langle \alpha
_i|=\sum_iP_{\alpha _i}\rho _{before}P_{\alpha _i}  \label{C.16"}
\end{equation}
where $p_i$ is the probability to measure $\alpha _i$ . Now if, after the
decoherence process, $\rho _{before}$ is a diagonal matrix, precisely $\rho
_{*}$ i.e. 
\begin{equation}
\rho _{before}=\sum_ip_i|\alpha _i\rangle \langle \alpha _i|  \label{C.16'''}
\end{equation}
and we only measure the observers in the CSCO $\{H,P_1,...P_n\},$ so the $%
P_{\alpha _i}$ are just the $P_\alpha =|\alpha \rangle \langle \alpha |,$ we
have: 
\begin{equation}
\rho _{after}=\sum_i|\alpha _i\rangle \langle \alpha _i|(\sum_jp_j|\beta
_j\rangle \langle \beta _j|)|\alpha _i\rangle \langle \alpha
_i|=\sum_ip_i|\alpha _i\rangle \langle \alpha _i|=\rho _{before}
\label{C.16""}
\end{equation}
So, in fact, the matrix $\rho _{*}$ is insensitive to the measurement of the
CSCO $\{H,P_1,...P_n\}$ $($ and also the CSCO $\{H,O_1,...,O_N\}$ where the
operators $O$ are related with the constants $m)$. This is the maximum
insensitivity we can get.

\subsection{Strong decoherence and records.}

If for any history $\overrightarrow{\alpha }$ there is a projector $R_\alpha 
$ such that $\{R_\alpha \}$ is not necessarily a complete set of projectors
in ${\cal H}$, in the sense that $\{R_\alpha |\psi \rangle \}$ is not
necessarily a basis of ${\cal H}$, and for any state $\rho $ it is: 
\begin{equation}
C_{\overrightarrow{\alpha }}\rho =R_\alpha \rho ,\text{ }R_\alpha R_\beta
=\delta _{\alpha \beta }P_\alpha  \label{7.1}
\end{equation}
we will say the we have {\it strong decoherence (}\cite{GyHI}, eq. (2.4))%
{\it .} As the $R_\alpha $ are timeless entities and as $C_{\overrightarrow{%
\alpha }}\rightarrow R_\alpha ,$ $R_\alpha $ can be considered as the $%
record $ of the history $\overrightarrow{\alpha }{\bf .}$ But $R_\alpha $
can also be considered as the record of not one but several decohered
histories, associated by unitary transformations \cite{GyHUCS}. So really $%
R_\alpha $ is the record of an equivalent class of histories.

It is clear that if these records exist we have medium decoherence. In fact: 
\begin{equation}
D(\overrightarrow{\alpha }{\bf ,}\overrightarrow{\alpha }^{\prime })=Tr(C_{%
\overrightarrow{\alpha }}^{\dagger }\rho C_{\overrightarrow{\alpha }^{\prime
}})=Tr(R_\alpha \rho R_{\alpha ^{\prime }})=Tr(\rho R_{\alpha ^{\prime
}}R_\alpha )=\delta _{\alpha \alpha ^{\prime }}p(\overrightarrow{\alpha }%
{\bf )}  \label{7.2}
\end{equation}
so strong decoherence implies medium decoherence.

In our case these final $R_\alpha $ exist and they are: 
\begin{equation}
R_\alpha =|\alpha \rangle \langle \alpha |=|xr\rangle \langle xr|=P_\alpha
\label{7.3}
\end{equation}
Thus the numbers $x,r_1,...,r_N$ can be considered as the record of the
corresponding final history.

\end{document}